\documentclass[aps,prc,twocolumn,superscriptaddress]{revtex4-1}

\usepackage{amsmath,amssymb,graphicx}
\usepackage{url}
\usepackage[colorlinks,urlcolor=blue]{hyperref}

\newcommand{\bfr}{\mathbf{r}}
\newcommand{\bfb}{\mathbf{b}}
\newcommand{\WK}[1]{{\color{blue} #1}}

\begin{document}

\title{Bayesian inference of the path-length dependence of jet energy loss}

\author{Jordan Wu}
\email{jordanwu4@berkeley.edu}
\affiliation{Nuclear Science Division, Lawrence Berkeley National Laboratory, Berkeley, California 94720, USA}
\affiliation{Physics Department, University of California, Berkeley, California 94720, USA}

\author{Weiyao Ke}
\email{weiyaoke@lanl.gov}
\affiliation{Nuclear Science Division, Lawrence Berkeley National Laboratory, Berkeley, California 94720, USA}
\affiliation{Physics Department, University of California, Berkeley, California 94720, USA}
\affiliation{Theoretical Division, Los Alamos National Laboratory, Los Alamos, NM 87545, USA}
\author{Xin-Nian Wang}
\email{xnwang@lbl.gov}
\affiliation{Nuclear Science Division, Lawrence Berkeley National Laboratory, Berkeley, California 94720, USA}
\affiliation{Physics Department, University of California, Berkeley, California 94720, USA}

\begin{abstract}
A simple model for medium modification of the jet function can be used to extract the jet energy loss distribution through a parameterized form. We carry out a comprehensive Bayesian analysis of the world data on single inclusive jet spectra in heavy-ion collisions at both RHIC and LHC energies. We extract the average jet energy loss $\langle \Delta E\rangle$ as a function of jet transverse momentum $p_T$ for each collision system and centrality independently. Assuming jet energy loss is proportional to the initial parton density $\rho \sim dN_{\rm ch}/d\eta/\pi R_{\rm eff}^2$ as estimated from the pseudorapidity density of charged hadron multiplicity $dN_{\rm ch}/d\eta$ and the effective system size $R_{\rm eff}\sim N_{\rm part}^{1/3}$ given by the number of participant nucleons $N_{\rm part}$, the scaled average jet energy loss $\langle \Delta E\rangle/\rho \sim R_{\rm eff}^{0.59} p_T^{0.13}\ln p_T $ for jet cone-size $R=0.4$ is found to have a momentum dependence that is slightly stronger than a logarithmic form while the system size or length dependence is slower than a linear one. The fluctuation of jet energy loss is, however, independent of the initial parton density or the system size. %These are consistent with results from Monte Carlo simulations of jet transport in a fast expanding quark-gluon plasma in high-energy heavy-ion collisions.
%Comparing the number of participants in the collision and the average energy loss reveals a size dependence for the energy loss of the jet, which can be seen across different collision energies when scaled by their charge multiplicities.
\end{abstract}



\maketitle

\section{Introduction}

Parton energy loss in dense medium was predicted to lead to the suppression of large transverse momentum hadrons and jets, known as jet quenching, in high-energy heavy-ion collisions~\cite{Gyulassy:1990ye,Wang:1992qdg}. Jet quenching was indeed observed in experiments at both the Relativistic Heavy-ion Collider (RHIC)~\cite{PHENIX:2001hpc,STAR:2002ggv,Wang:2004dn} and Large Hadron Collider (LHC)~\cite{ATLAS:2010isq,ALICE:2010yje,CMS:2012aa,CMS:2012aa,CMS:2016uxf,ALICE:2019qyj}. Phenomenological studies have extracted the jet transport coefficient from comparisons between experimental data and model calculations~\cite{Chen:2010te,JET:2013cls,JETSCAPE:2021ehl,Liu:2021dpm,Apolinario:2022vzg,Xie:2022ght,Xie:2022fak} whose values point to the formation of the quark-gluon plasma with extremely high temperatures in high-energy heavy-ion collisions. These data-model comparisons are all based on perturbative QCD calculations of collisional and radiative parton energy loss inside QGP. In a static and uniform QGP medium, the total radiative parton energy loss is predicted to have a quadratic path-length dependence because of the non-Abelian Landau-Pomeranchuck-Migdal interference in gluon radiation induced by multiple scattering~\cite{Baier:1996kr,Zakharov:1996fv,Wiedemann:2000za,Gyulassy:2000fs,Guo:2000nz,Wang:2001ifa,Majumder:2009ge}. It is also proportional to the jet transport coefficient, which is proportional to the local %parton
\WK{color charge} density. Taking into account the rapid longitudinal and transverse expansion in high-energy heavy-ion collisions, the effective total parton energy loss averaged over the azimuthal angle is approximately linear in the system size or the average escape time~\cite{Gyulassy:2001kr}.  Such a system size dependence is consistent with the effective parton energy loss extracted from the measured suppression of single inclusive hadrons~\cite{Arleo:2022shs}.

One can define the effective energy loss for a reconstructed jet with a given jet cone size $R$ as the difference between the energy of a jet in proton+proton ($p$+$p$) collisions and the energy of the final jet in heavy-ion ($A$+$A$) collisions that originates from the same hard process in $p$+$p$ collisions. The relation between the jet energy loss and that of an individual parton is not straightforward since some of the radiated gluons can end up inside the jet cone as part of the final jet. In addition, some of the lost energy carried by the recoil medium partons as part of the jet-induced medium response can also contribute to the final jet energy with a given jet cone size $R$. The momentum and system size dependence of the jet energy loss can only be studied through calculations that take into account of both the above effects~\cite{He:2015pra,Wang:2016fds,He:2018xjv}. 

Since the jet production cross section can be factorized as the convolution of the hard parton cross section and parton jet function which can be further expressed as a convolution of jet function in vacuum (in $p$+$p$ collisions) and jet energy loss distribution, one can extract the jet energy loss distribution from the experimental data with Bayesian inference~\cite{He:2018gks}. In this study we will follow the same procedure of Bayesian inference in Ref.~\cite{He:2018gks} that extracted the momentum dependence of the jet energy loss in the most central Pb+Pb collisions at LHC. We carry out a systematic analysis of world data on the single inclusive jet cross sections in $p$+$p$ and $A$+$A$ collisions with all possible centrality selections at both RHIC and LHC energies. We will focus on the system size dependence of the extracted jet energy loss as well as its scaling behavior with respect to the charged hadron pseudorapidity density in the final state.

\section{Jet production cross section}

We first briefly summarize here the pQCD parton model for jet production that we use to extract jet energy loss distributions from experimental data. The differential cross section for single inclusive jet production in $p$+$p$ collisions can be expressed in a factorized and schematical form~\cite{Kang:2016mcy,Kang:2017frl},
\begin{equation}
\frac{d\sigma_{pp}^{\rm jet}}{dp_Td\eta}=\sum_{a,b,c}\int f_{a/{\rm p}}\otimes f_{b/{\rm p}}\otimes H_{ab}^{c}\otimes J_c(p_T, R|p_{Tc}),
\end{equation}
where $f_{a/p}$ is the parton distribution function of proton, $H_{ab}^c$ is the hard function for parton scattering $a+b\rightarrow c+X$ and $J_c(p_T, R|p_{Tc})$ is the semi-inclusive jet function that describes the probability for a parent parton $c$ with initial transverse momentum $p_{Tc}$ to produce a jet with transverse energy $p_T$ and jet-cone size $R$.  Similarly, the single inclusive jet production cross section in $A$+$A$ collisions can be written as,
\begin{eqnarray}
\frac{d \sigma^{\rm jet}_{AA}}{dp_{T}d\eta} & = &\sum_{a,b,c}  \int d^2\bfr d^2\bfb  t_A(\bfr) t_A(|\bfb-\bfr|) \frac{d\phi_c}{2\pi}   \nonumber\\
&& \hspace{-0.4in} \times f_{a/{A}} \otimes f_{b/{A}} \otimes H_{ab}^c \otimes  \widetilde{J}_{c}(p_T,R,\bfr,\bfb,\phi_c|p_{Tc}),
\label{eq:cs.aa}
\end{eqnarray}
where $t_{A}(r)$ is the nuclear thickness function with normalization $\int d^2r t_{A}(r)=A$, $f_{a/A}$ is the parton distribution function per nucleon inside the nucleus $A$, $\bfr$ is the transverse coordinate of the hard production vertex of the initial parton ($c$), $\phi_c$ is azimuthal angle between its transverse momentum $p_{Tc}$ and the impact parameter $\bfb$ of the nucleus-nucleus collision, and $\widetilde{J}_{c}(p_T,R,\bfr,\bfb,\phi_c|p_{Tc})$ is the medium-modified semi-inclusive jet function for a given path of jet propagation in the QGP. The impact parameter $\bfb$ is integrated over a range that is determined by the centrality class of the nucleus-nucleus collisions according to experimental measurements. 

The modified jet functions $\widetilde{J}_{c}$ take into account the jet energy loss due to induced gluon radiation and collisional energy loss carried by the medium response outside the jet cone. Assuming the in-medium jet function can be approximated by a shift in the jet energy of the vacuum jet function and further considering event-by-event fluctuates of jet energy loss for a given propagation path, the medium-modified jet function can be expressed as the convolution, 
\begin{eqnarray}
 \widetilde{J}_{c}(p_T,R,\bfr,\bfb,\phi_c|p_{Tc}) &=&\int_0^\infty d\Delta E  J_c(p_T+\Delta E, R|p_{Tc}) \nonumber \\
&&\hspace{-0.5in} \times w_c(\Delta E, p_{T}+\Delta E, R, \bfr,\bfb,\phi_c),
\end{eqnarray}
of the jet function in vacuum with transverse energy $p_T+\Delta E$ and a jet energy loss distribution $w_c$ for a given path specified by $\bfr$, $\bfb$ and $\phi_c$. The jet transverse energy loss is defined as the difference between the jet transverse energy in $p$+$p$ and $A$+$A$ collisions originating from the same initial parton $c$. 

One can define the jet energy loss distribution averaged over the initial parton production point and propagation direction,
\begin{eqnarray}
W^{c}_{AA}(\Delta E, p_{T}, R)&=& \int d^2\bfr d^2\bfb  \frac{t_A(r) t_A(|\bfb-\bfr|)}{N_{\rm bin}(b)} \nonumber \\
&\times& \int  \frac{d\phi_c}{2\pi} w_c(\Delta E, p_{T}, R, \bfr,\bfb,\phi_c),~~
\end{eqnarray}
for a given centrality class of $A$+$A$ collisions, where $N_{\rm bin}(b)=\int d^2\bfr d^2\bfb  t_A(r) t_A(|\bfb-\bfr|)$ is the number of binary collisions. 
Then, the single inclusive jet production cross section in $A$+$A$ collision can be written as
\begin{eqnarray}
\frac{d \sigma^{\rm jet}_{AA}}{dp_{T}d\eta} & = &N_{\rm bin}(b)\sum_{a,b,c}  \int  d\Delta E W^{c}_{AA}(\Delta E, p_{T}+\Delta E,R) \nonumber\\
&& \hspace{-0.2in} \times f_{a/{A}} \otimes f_{b/{A}} \otimes H_{ab}^c \otimes  J_{c}(p_T+\Delta E,R|p_{Tc}),
\label{eq:cs.aa2}
\end{eqnarray}
In principle, the nuclear modification of the parton distributions (nPDF) should be considered in the above jet production cross section \cite{Adhya:2021kws}. However, the nuclear modification of PDF is usually limited to small $x$ or very large $x$ (for EMC effect) and low $Q^2$ regions~\cite{Eskola:2016oht}. For our analyses in this study, we will limit the jet transverse momentum to $15<p_T<30$ GeV/$c$ at RHIC and $50<p_T<800$ GeV/$c$ at LHC which correspond to $0.15<x\approx 2p_T/\sqrt{s}<0.3$ and $0.02<x<0.32$ in the central rapidity region ($y=0$), respectively. For such large $Q^2\approx p_T^2$, the nuclear modification of the PDF is mostly negligible in these regions of $x$~\cite{Eskola:2016oht}. This is also confirmed by NLO pQCD parton model calculations~\cite{Xie:2020zdb} and indicated by recent experimental data on the single inclusive jet cross section in minimum-bias p+Pb collisions at LHC~\cite{ATLAS:2014cpa}. Furthermore, the jet cross sections also do not depend on the isospin of the initial quark flavors.

Under the assumption that jet cross section in the considered kinematic region is not sensitive to the nuclear modification of parton distribution,
the single inclusive jet cross section in $A$+$A$ collisions can be expressed as the convolution of jet cross section in $p$+$p$ collisions and a flavor-averaged (quarks and gluon) jet energy loss distribution $W_{AA}$. The suppression factor for single inclusive jet production in $A$+$A$ collisions can be written as
\begin{eqnarray}
 R_{AA}(p_{T}) &\approx& \frac{1}{d\sigma^{\rm jet}_{pp}(p_T)} \int d\Delta E d\sigma^{\rm jet}_{pp}(p_{T} + \Delta E) \nonumber\\
&\times& W_{AA}(\Delta E, p_T+\Delta E, R).
 \label{eq:raa}
\end{eqnarray}
This formula has been used  for the study of jet suppression~\cite{Spousta:2015fca,Mehtar-Tani:2017web} and similar approximate expression for single inclusive hadron spectra has been used in Refs.~\cite{Arleo:2017ntr,Arleo:2022shs,Baier:2001yt} assuming a constant average momentum fraction of hadrons $z_h=p_{Th}/p_T$ in the energy loss distribution. This is also the expression we use to extract jet energy loss distribution from experimental data through Bayesian inference in this study. 
In this study, we will use Pythia8~\cite{Sjostrand:2007gs} simulations to calculate the jet production cross sections in $p+p$ collisions which are shown to describe the experimental data very well at both RHIC and LHC energies.

In the above equation the jet energy loss distribution is averaged over parton flavors (quark versus gluon) weighted by their respective cross sections. It is possible to extend this analysis to include the flavor dependence with given fractions of the total jet cross section by the pQCD parton model calculations~\cite{Pablos:2022mrx,Zhang:2023oid,Zhang:2022rby}. This will double the number of parameters in the jet energy loss distributions. Investigation of such flavor-dependent Bayesian inference and the corresponding constraints on the momentum and path-length dependence from single inclusive jet modification is beyond the scope of this study and will be left for future studies.

%Here, we run Pythia8 simulations  at $\sqrt{s}=200$ GeV, $2.76$ TeV, and $5.02$ TeV to obtain $d\sigma_{pp}^{\rm jet}/dp_T$ ($R=0.4$)  used in Eq. \ref{eq:raa}.

%\WK{(Do we need to comment on the quark-vs-gluon jet fraction. Such as the kinematics region that we considered has an approximate constant ratio of quark/gluon jets and existing model calculations suggests a moderate difference in quark versus gluon jet energy loss. So we are not going to differentiate the flavor dependence of the in-medium jet function)}

\section{Bayesian analysis of single jet suppression $R_{\rm AA}$}

Using Eq.~(\ref{eq:raa}) that relates the nuclear modification factor $R_{AA}$ of single inclusive jet cross section and the jet energy loss distribution $W_{AA}$, we attempt to reverse engineer $W_{AA}$ from experimental data on $R_{AA}$. Though an exact inversion of the convolution problem can be ill-defined, one can determine the probability distribution of $W_{AA}$ using Bayesian inference. This method has found a broad application in heavy-ion collisions, including the extraction of the QCD equation of state at high-temperature~\cite{Pratt:2015zsa}, the QGP shear and bulk viscosity~\cite{Bernhard:2016tnd,Bernhard:2019bmu,JETSCAPE:2020mzn,JETSCAPE:2020shq,Nijs:2020roc,Nijs:2020ors,Parkkila:2021tqq,Parkkila:2021yha}, the heavy quark diffusion constant~\cite{Xu:2017obm,Ke:2018tsh,Liu:2021dpm}, and the jet transport coefficient in the QGP~\cite{Ke:2020clc,JETSCAPE:2021ehl,Liu:2021dpm,Xie:2022ght,Xie:2022fak}. 
We will follow Ref.~\cite{He:2018gks} and carry out a comprehensive Bayesian analysis of the world data on single inclusive jet spectra in heavy-ion collisions with different centralities at both RHIC and LHC energies and extract the colliding energy, jet momentum and system size dependence of the jet energy loss and its fluctuations.

We assume that the average jet energy loss $\langle\Delta E\rangle_i$ in a given colliding system and centrality (collectively labeled by $i$) is given by
\begin{eqnarray}
\langle\Delta E\rangle_i = \beta_i \left(\frac{p_{T}}{p_{T\rm ref}}\right)^{\gamma_i} \ln\left(\frac{p_{T}}{p_{T\rm ref}}\right), \\\nonumber
\end{eqnarray}
with $p_{T\rm ref}=1$ GeV/$c$ as a reference momentum. %The power-law part of the function is to 
$\gamma_i>0$ guarantees that the jet energy loss vanishes when jet momentum goes to zero. The jet energy loss distribution is assumed to only depend on the self-normalized energy loss fluctuation $x=\Delta E/\langle \Delta E\rangle$, i.e., $W_{AA}(\Delta E, p_{T}, R)\approx W_{AA}(x, R)$. Such an approximate feature was corroborated in the previously study using the linear Boltzmann transport (LBT) model simulations~\cite{He:2018xjv}.
We parametrize the energy loss fluctuation as 
\begin{equation}
W_{AA}(x)=\frac{\alpha_i^{\alpha_i} x^{\alpha_i-1}e^{-\alpha_i x}}{\Gamma(\alpha_i)}.
\label{eq:waa}
\end{equation}
From now on, we will omit the label of the jet cone size $R$ dependence of $W_{AA}$ as we will only use jet measurements  with $R=0.4$ for the rest of this study.

By systematic comparing to dataset $D_i$ with $N_i$ data points, the posterior probability distributions of the parameters $\theta_i\equiv[\alpha_i,\beta_i,\gamma_i]$ are given by the Bayesian theorem, 
\begin{equation}
P(\theta_i |D_i) = \frac{P(D_i| \theta_i) P(\theta_i)}{P(D_i)},
\label{eq:bayesian}
\end{equation}
 where $P(\theta_i | D_i)$ is the posterior distribution after model-data comparison.
 $P(\theta_i)$ is the prior distribution of model parameters. $P(D_i | \theta_i)$ is the likelihood function between experimental data
 and model $(M)$ calculations using parameter $\theta_i$. The likelihood is assumed to take a Gaussian form,
 \begin{eqnarray}
\ln P(D_i | \theta_i) &=& -\frac{N_i}{2}\ln(2\pi) -\frac{1}{2}\ln|\Sigma_i| \nonumber \\
&-& \frac{1}{2}[M(\theta_i)-D_i]^T \Sigma^{-1}[M(\theta_i)-D_i].~~
 \end{eqnarray}
$M(\theta_i)-D_i$ is the discrepancy vector between model calculation and data. The covariance matrix $\Sigma_i$ contains experimental and computational uncertainty.
Uncertainties of the experimental data are assumed to be uncorrelated between different data points. 
The normalization $P(D_i) = \int d\theta_i  P(D_i | \theta_i) P(\theta_i)$ is called the evidence. 
The properties of the posterior distribution can be explored with importance sampling. 
The common practice is the Markov Chain Monte Carlo (MCMC)~\cite{Andrieu2003} method that performs importance sampling $\theta_i$ according to $P(D_i|\theta_i)P(\theta_i)$. Here, we use the affine-invariant MCMC algorithm~\cite{2010CAMCS...5...65G} as implemented in the \texttt{emcee} package~\cite{Foreman_Mackey_2013}.
Projecting the posterior samples to lower dimensions is equivalent to the marginalization of the high-dimensional distribution. 
From the marginalization procedure one can define the one-parameter posterior distributions and the pairwise correlations. However, the physically meaningful quantify is not individual parameter but the functional form of the average energy loss. We can marginalized all three parameters to obtain the posterior distribution of the averaged energy loss functional as
\begin{eqnarray}
P[\langle\Delta E_i\rangle(p_T)] = \int \langle\Delta E\rangle(p_T; \beta_i, \gamma_i) P(\theta_i|D_i) d\theta_i.
\end{eqnarray}
Then, one can define the median, and percentile credible interval (C.I.) of the posterior energy loss at each jet energy.

It is important to note that we do not impose that the parameters $[\alpha_i,\beta_i,\gamma_i]$ to be the same for different colliding systems (colliding energy and centrality).  They will be extracted independently for each experimental data set ``$i$''. We will then analyze these ``piecewise'' information to study the momentum dependence extracted from different colliding systems and check whether they are consistent with each other. Eventually, we will determine the path-length dependence of the jet energy loss by correlating the extracted jet energy loss with the averaged path-length for each centrality class of collisions at each colliding energy.

 \section{Extract energy loss parameters}

Using the above Bayesian method we have carried out analyses of experimental data on the nuclear modification of single inclusive jet spectra and extract the jet energy loss distributions in Pb+Pb collisions at $\sqrt{s}=5.02$ and 2.76 TeV and Au+Au collisions at $\sqrt{s}=200$ GeV for all available centrality classes.  The experimental data on single inclusive jet spectra are from ATLAS~\cite{ATLAS:2014ipv,ATLAS:2018gwx} for Pb+Pb collisions at $\sqrt{s}=5.02$ and 2.76 TeV with 8 centrality classes (0-10\%,10-20\%,20-30\%,30-40\%, 40-50\%, 50-60\%, 60-70\% and 70-80\%),  ALICE~\cite{ALICE:2019qyj} for Pb+Pb collisions at $\sqrt{s}=5.02$ TeV with 0-10\% centrality class and STAR~\cite{STAR:2020xiv} for Au+Au collisions at $\sqrt{s}=200$ GeV with 2 centrality class (0-10\%, 60-80\%).
These experimental measurements are all at the central rapidity region with jet cone size $R=0.4$.
In Tab. \ref{tab:expdata}, we list these dataset along with the range of transverse momentum of jet measurements and the value of $\langle N_{\rm part}\rangle^{1/3}$ which is proportional to the system size or the averaged path length for a given centrality class.
Values of $\langle N_{\rm part}\rangle$ are obtained from the Glauber model that are used in experimental analysis \cite{ALICE:2013hur,ALICE:2015juo,PHENIX:2015tbb}. 
We have not included the jet data from CMS experiment~\cite{CMS:2016uxf}. Nevertheless, given the relatively larger experimental uncertainty of CMS data than that from ATLAS in a similar range of the jet transverse momentum $p_T$, we do not expect a significant change of accuracy for the present Bayesian analysis.

\begin{table}[ht!]
    \centering
    \begin{tabular}{c|c|c|c|c}
    \hline
    System & Centrality & $p_T$ range [GeV/$c$] & Refs. & $\langle N_{\rm part}\rangle^{1/3}$\\
    \hline
     Pb+Pb  &  0-10\% & $(100, 1000)$ & ATLAS~\cite{ATLAS:2018gwx} & 7.11\\
    5.02 TeV       &  10-20\% & $(100, 630)$  & & 6.41\\
           &  20-30\% & $(79, 630)$ & & 5.73\\
           &  30-40\% & $(79, 630)$ & & 5.08\\
           &  40-50\% & $(50, 398)$ & & 4.42\\
           &  50-60\% & $(50, 398)$ & & 3.77\\
           &  60-70\% & $(50, 398)$ & & 3.12\\
           &  70-80\% & $(50, 251)$ & & 2.50\\
     &  0-10\% & $(60,140)$  & ALICE~\cite{ALICE:2019qyj} & 7.11\\
     \hline
     Pb+Pb  &  0-10\% & $(50, 398)$ & ATLAS~\cite{ATLAS:2014ipv} & 7.09\\
     2.76 TeV      &  10-20\% & $(50, 316)$ & & 6.38\\
           &  20-30\% & $(50, 316))$ & & 5.71\\
           &  30-40\% & $(39, 316)$ & & 5.05\\
           &  40-50\% & $(39, 316)$ & & 4.39\\
           &  50-60\% & $(39, 316)$ & & 3.61\\
           &  60-70\% & $(39, 251)$ & & 3.10\\
           &  70-80\% & $(39, 199)$ & & 2.48\\
    \hline
     Au+Au  &  0-10\% & $(15.57, 29.14)$  & STAR~\cite{STAR:2020xiv} & 6.86\\
     200 GeV       &  60-80\% & $(14, 24.52)$ & & 2.62\\
    \hline
    \end{tabular}
    \caption{List of experimental data used in this analysis. }
    \label{tab:expdata}
\end{table}

\begin{figure}[ht!]
\includegraphics[width=\columnwidth]{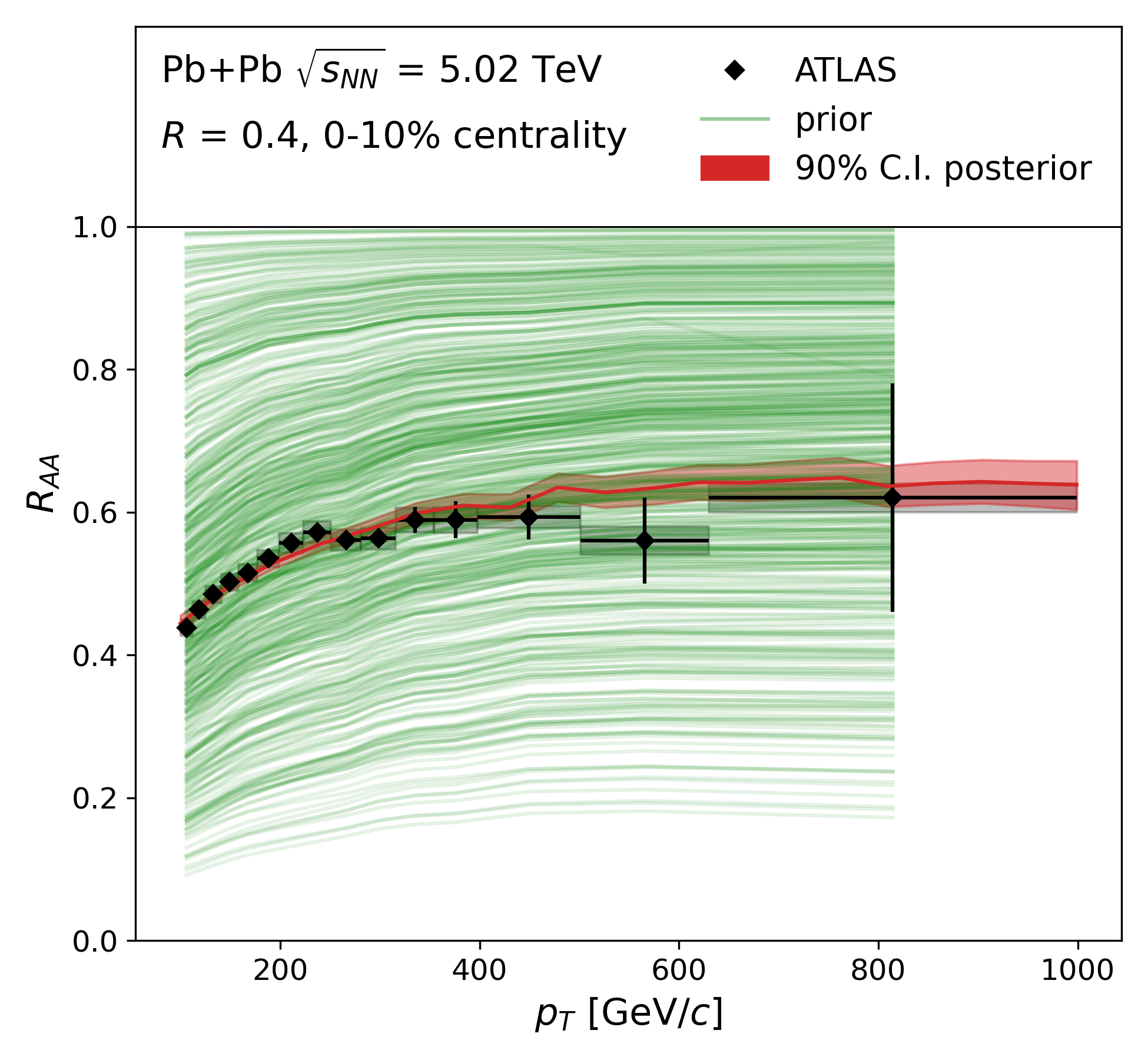}
%\vspace{-0.15in}
 \caption{
 A demonstration of calibration to the ATLAS data on single inclusive jet $R_{AA}$ in 0-10\% Pb+Pb collisions at $\sqrt{s}=5.02$ TeV~\cite{ATLAS:2018gwx}.
$R_{AA}$ calculated with the prior samples of the parameters  $[\alpha, \beta, \gamma]$ for the jet energy loss distribution are shown as green lines. The posterior of $R_{AA}$ at 90\% credible interval (C.I.) after the calibration is shown as the red band. The same procedure has been performed for each dataset listed in Tab. \ref{tab:expdata}.
 }
 \label{fig:postRAA_ATLAS_5020_0-10}
\end{figure}

\begin{figure}[ht!]
\centering
\includegraphics[width=\columnwidth]{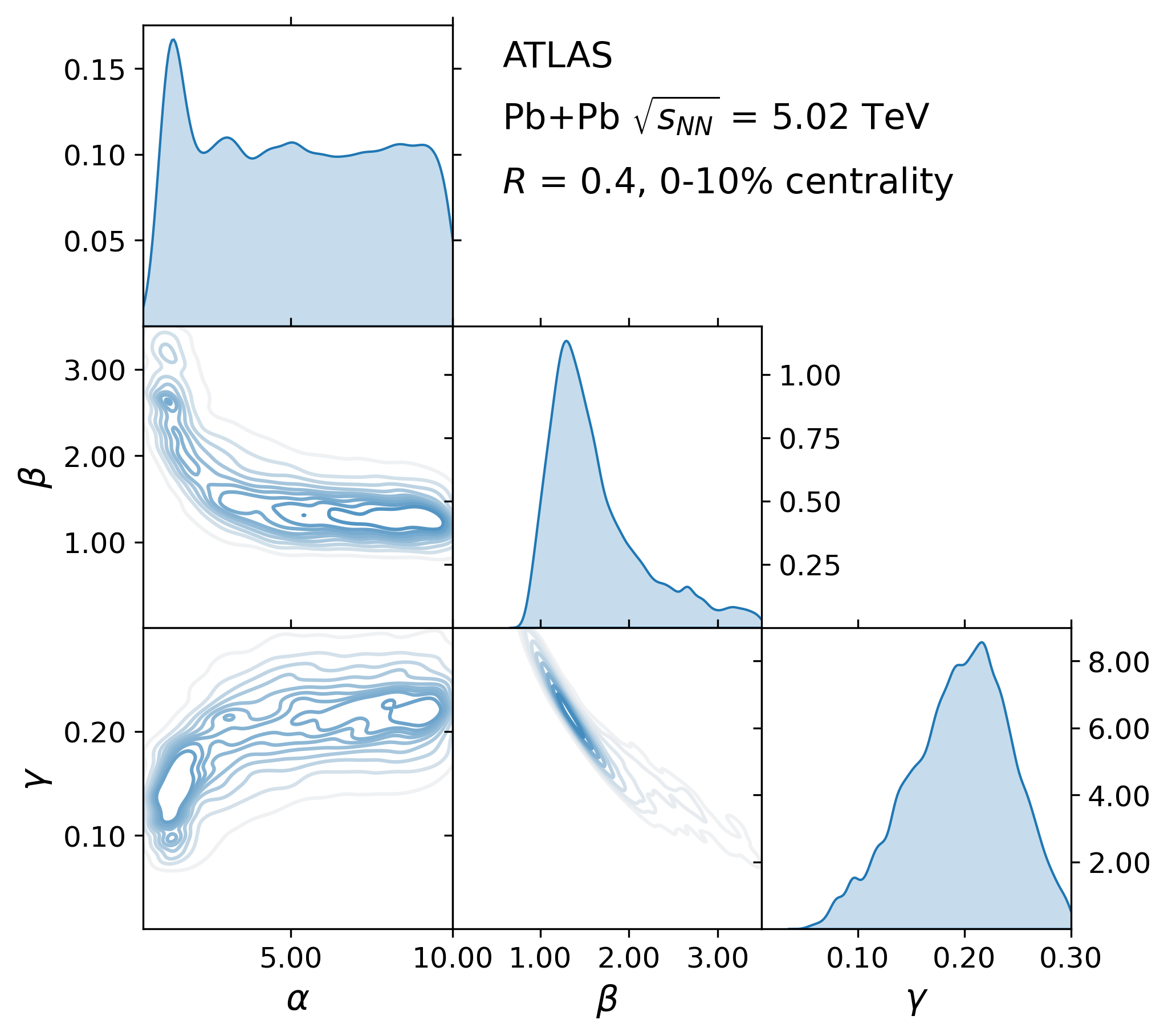}
 \caption{The posterior distribution of $[\alpha, \beta, \gamma]$ after calibrating to the ATLAS data on single inclusive jet $R_{AA}$ in 0-10\% Pb+Pb collisions at $\sqrt{s}=5.02$ TeV~\cite{ATLAS:2018gwx}. The same procedure has been performed for each dataset listed in Tab. \ref{tab:expdata}}
 \label{fig:postparams_ATLAS_5020_0-10}
\end{figure}

\begin{figure}[ht!]
\centering
\includegraphics[width=\columnwidth]{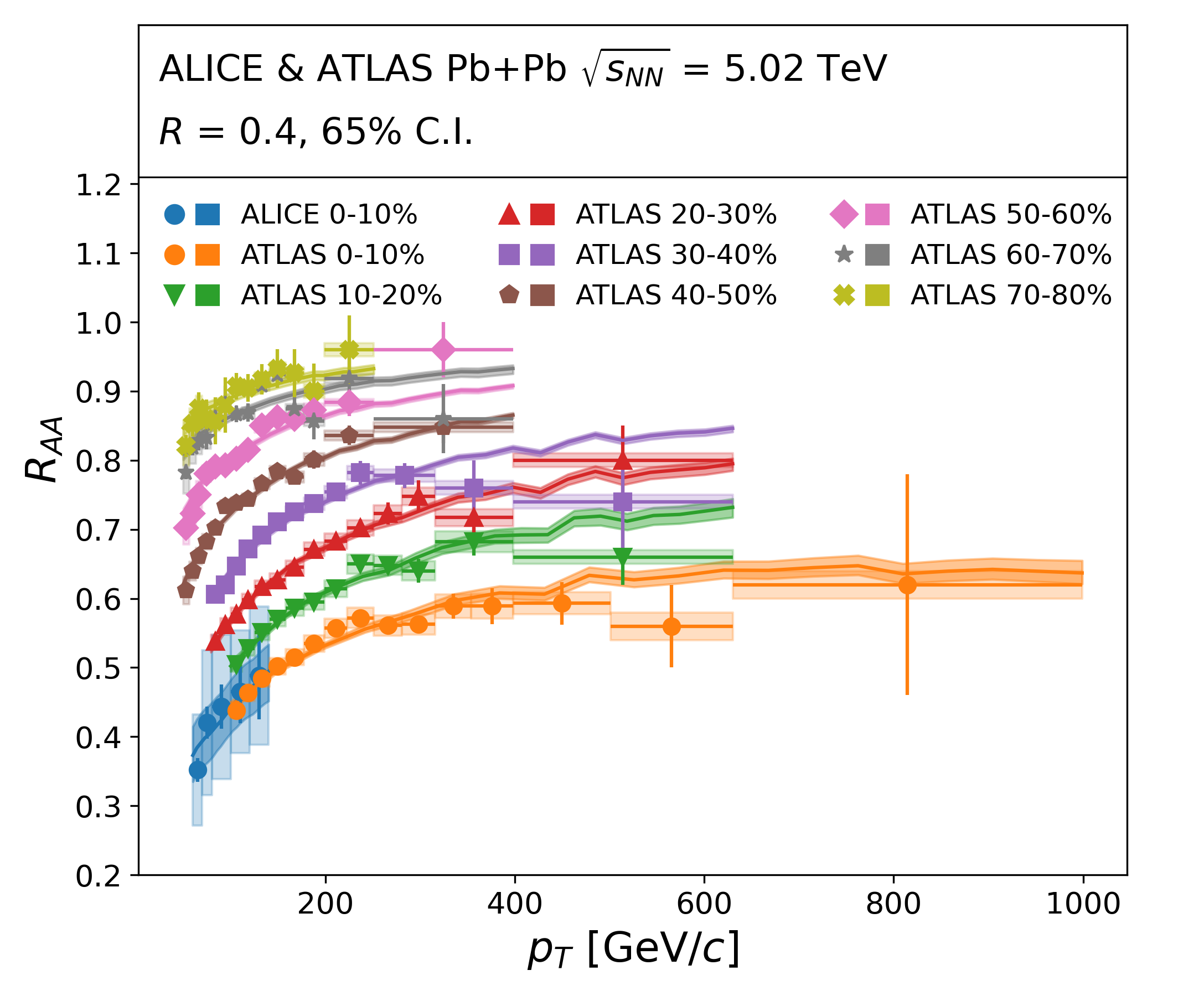}
\caption{Posterior predictions with median (colored lines) and 65\% credible intervals (shaded bands) compared to the experimental data~\cite{ATLAS:2018gwx,ALICE:2019qyj} on single inclusive jet ($R=0.4$) $R_{AA}$ in  Pb+Pb collisions at $\sqrt{s}=5.02$ TeV with different centralities.
}
\label{fig:postRAA_ATLAS_5020}
\end{figure}

\begin{figure}[ht!]
\centering
\includegraphics[width=1.1\columnwidth]{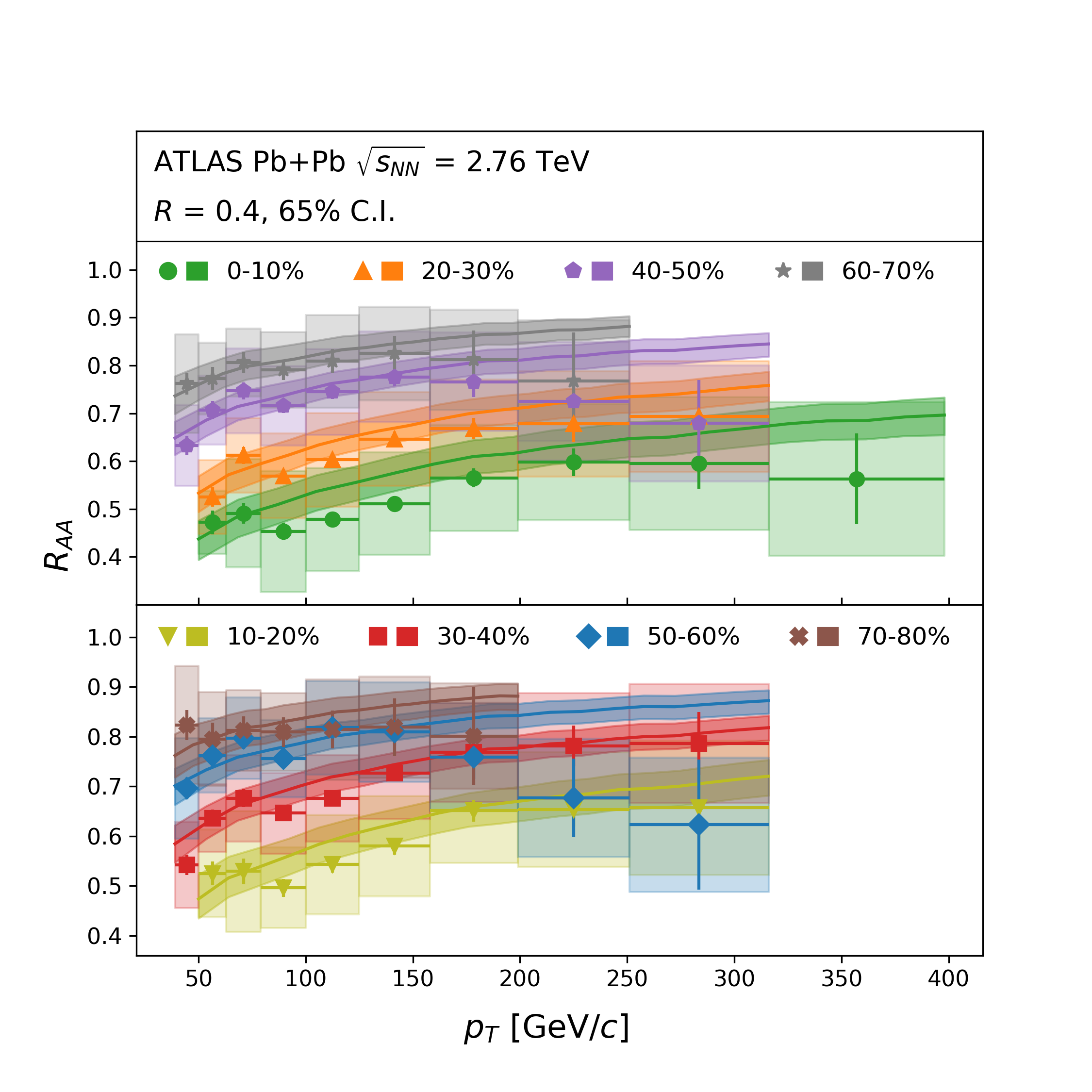}
 \caption{The same as Fig.~\ref{fig:postRAA_ATLAS_5020} except for Pb+Pb collisions at $\sqrt{s}=2.76$ TeV with data from Ref.\cite{ATLAS:2014ipv}.
 }
 \label{fig:postRAA_ATLAS_2760}
\end{figure}

\begin{figure}[ht!]
\centering
\includegraphics[width=\columnwidth]{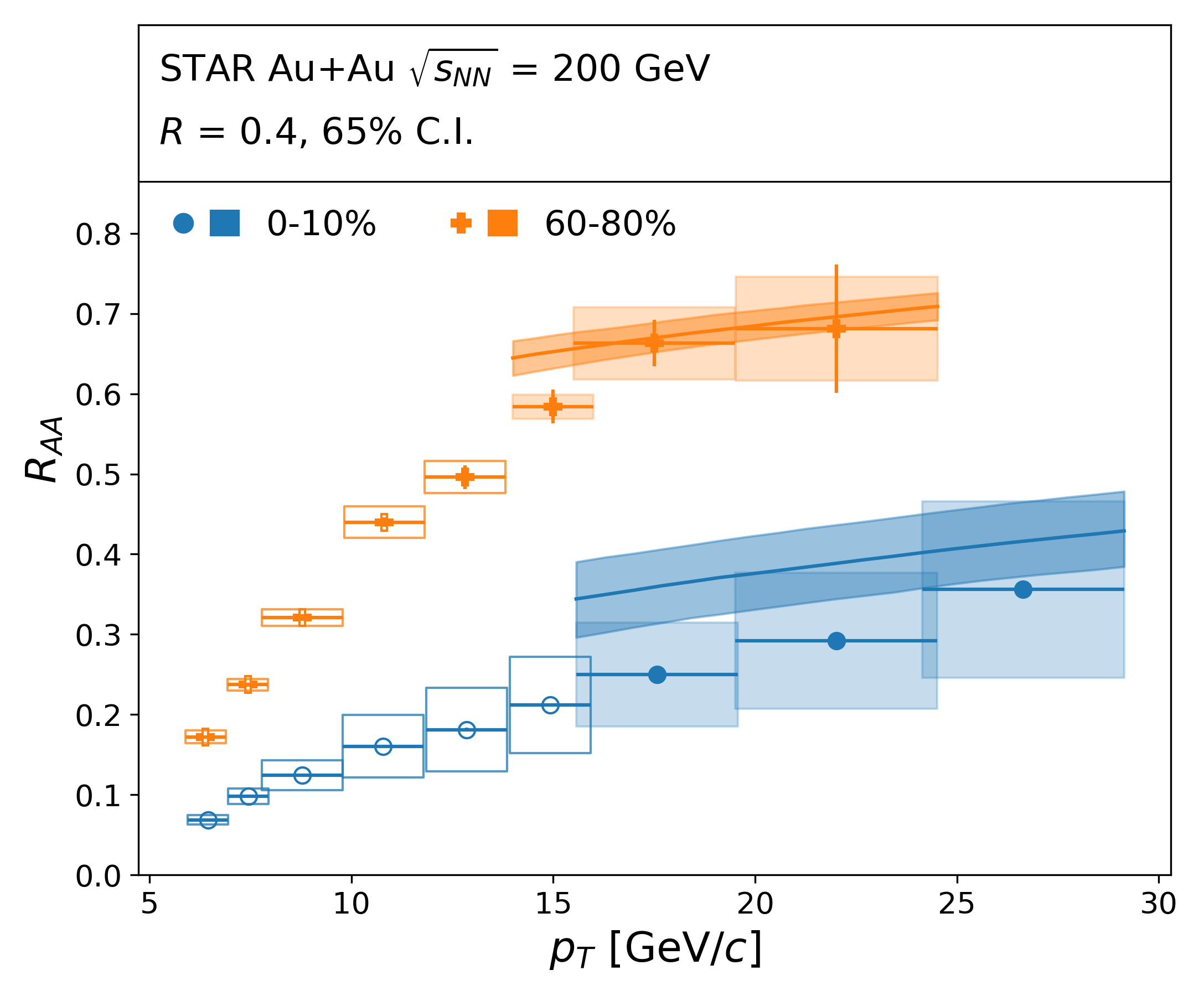}
 \caption{The same as Fig.~\ref{fig:postRAA_ATLAS_5020} except for Au+Au collisions at $\sqrt{s}=200$ GeV with data from Ref.\cite{STAR:2020xiv}.
 }
 \label{fig:postRAA_STAR_200}
\end{figure}

\begin{figure}[ht!]
\centering
\includegraphics[width=\columnwidth]{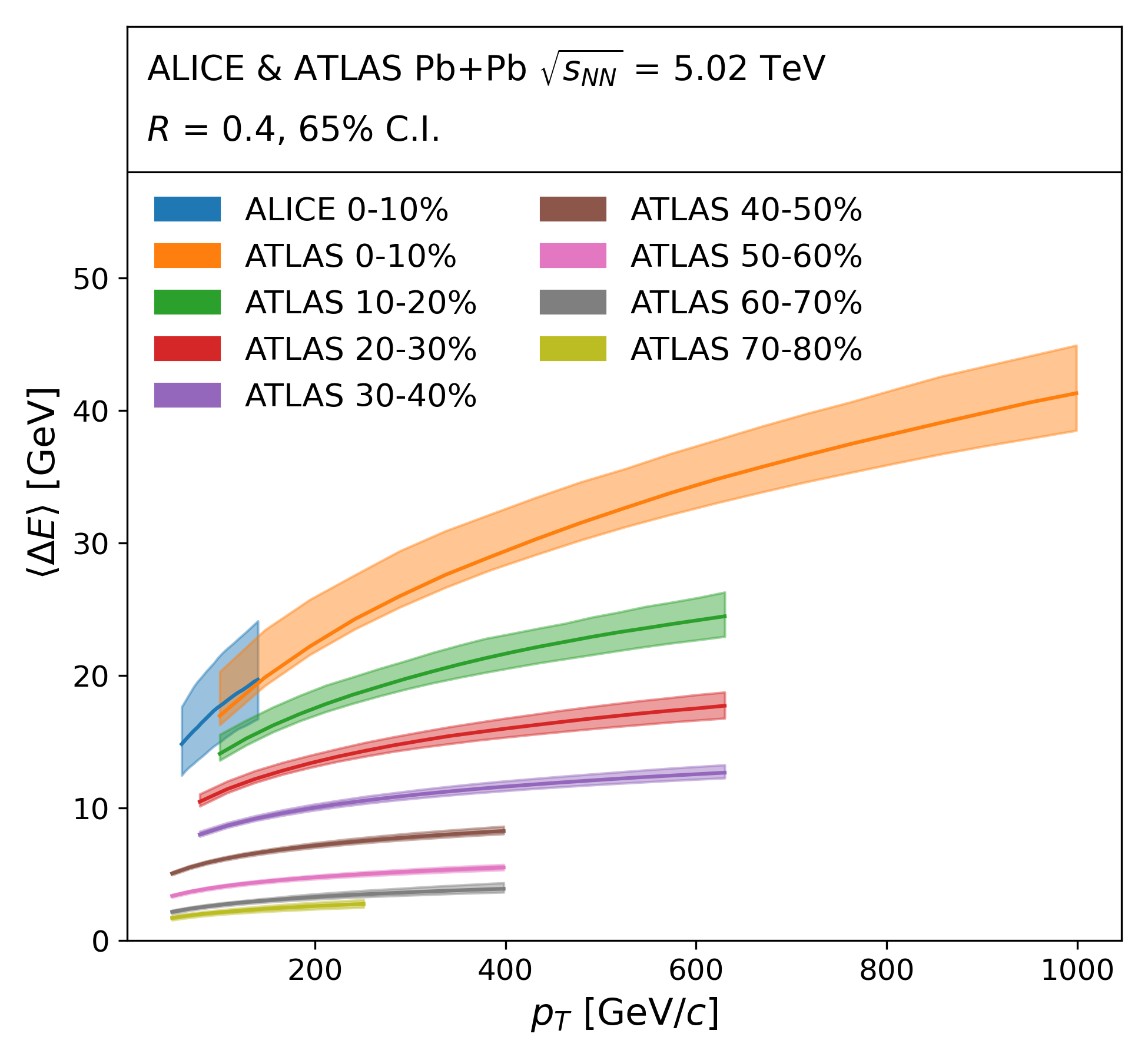}
 \caption{The posterior of the average jet energy loss ($R=0.4$) with the median (solid lines) with 65\% C.I. (solid bands) as a function of the initial jet transverse momentum $p_T$ extracted from experimental data for Pb+Pb collisions at $\sqrt{s}=5.02$ TeV with different centralities.}
 \label{fig:postElossATLAS_5020}
\end{figure}

\begin{figure}[ht!]
\centering
\includegraphics[width=\columnwidth]{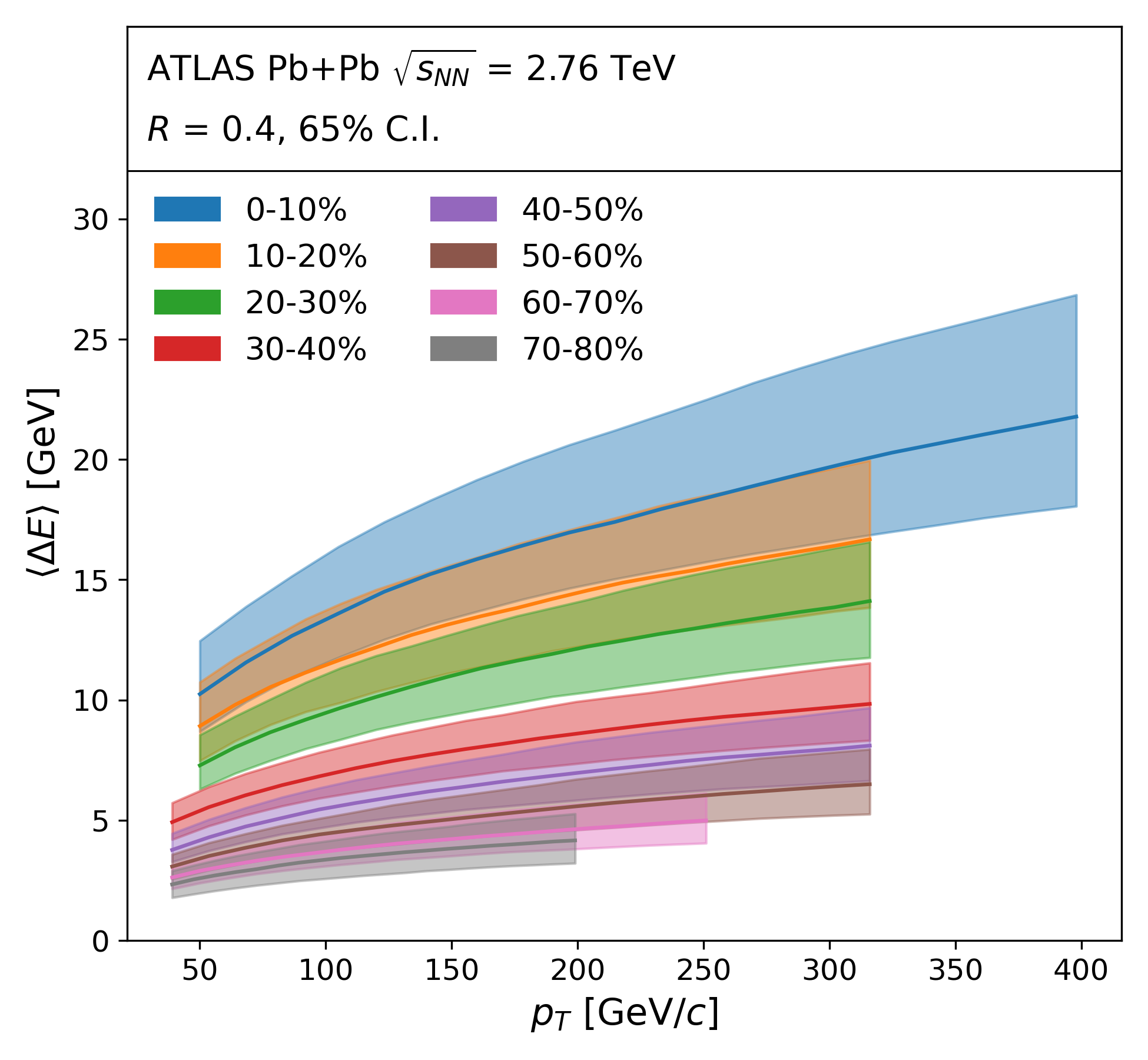}
 \caption{ The same as Fig.~\ref{fig:postElossATLAS_5020}
   except for Pb+Pb collisions at $\sqrt{s}=2.76$ TeV.
 }
 \label{fig:postElossATLAS_2760}
\end{figure}

\begin{figure}[ht!]
\centering
\includegraphics[width=\columnwidth]{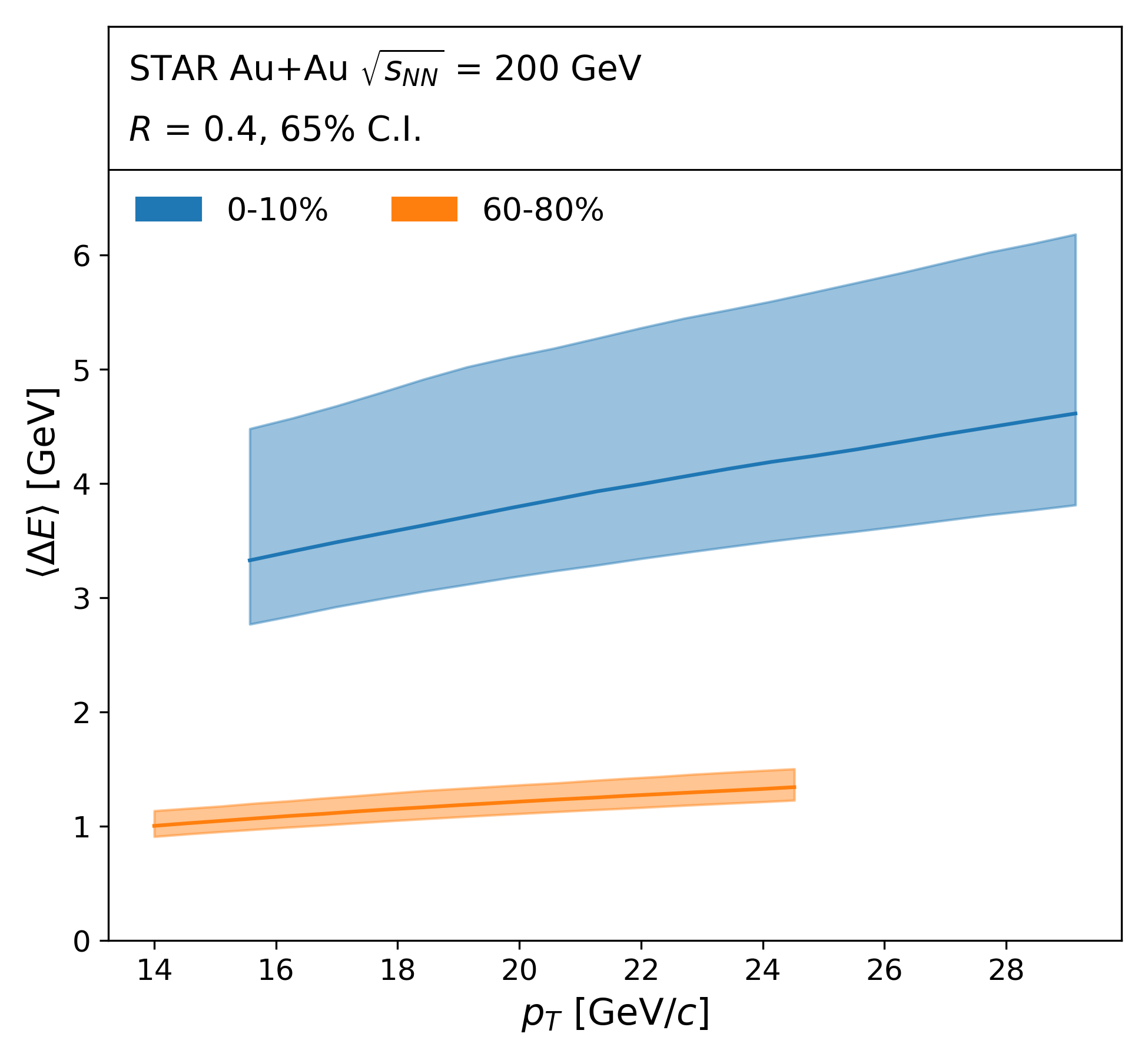}
 \caption{ The same as Fig.~\ref{fig:postElossATLAS_5020}
   except for Au+Au collisions at $\sqrt{s}=200$ GeV.
 }
 \label{fig:postElossSTAR_200}
\end{figure}
 
In Figs.~\ref{fig:postRAA_ATLAS_5020_0-10} and~\ref{fig:postparams_ATLAS_5020_0-10}, we show the prior and posterior observables and posterior distribution of parameters by calibrating to the measured  single jet $R_{AA}$ in $0-10$\% Pb+Pb collisions at $\sqrt{s}=5.02$ TeV from the ATLAS Collaboration~\cite{ATLAS:2018gwx}. 
The prior ranges of parameters are $0<\alpha<10$, $0<\beta<3.5$, and $0<\gamma<0.3$, which provide a good prior coverage (green lines) of the experimental data. Calculations using the posterior distributions of parameters at 90\% credible interval (C.I.) (solid red band) are in good agreement with the experimental data. 

Similarly, independent calibrations have been performed for each dataset in Tab. \ref{tab:expdata}. Shown in Figs.~\ref{fig:postRAA_ATLAS_5020}, \ref{fig:postRAA_ATLAS_2760} and \ref{fig:postRAA_STAR_200} are the final fits to the single inclusive jet suppression factor $R_{AA}$ as functions of the final jet transverse momentum $p_T$ in $A$+$A$ collisions with different centrality classes at three different colliding energies. The solid lines are the mean averages of the fit and shaded bands are uncertainties at 65\% C.I.
We observe that errors at 65\% C.I. are consistent with the experimental uncertainties.
Note that the STAR experiment measures charged jets and uses Pythia8 simulations as the $p+p$ baseline, while ALICE and ATLAS measures full jets with experimental data as the $p+p$ baseline. 
Therefore, we have scaled the jet transverse momentum by 2/3 when Eq.~(\ref{eq:raa}) is used to fit the STAR data on charged jets.
Furthermore, STAR charged jets covers relatively low $p_T^{\rm jet}$ region, a leading charged particle trigger $p_T^{h^{\pm}}>5$ GeV/$c$ is used to suppress ``fake jet'' contributions. However, the leading particle trigger also biases the jet suppression. Therefore, we will only calibrate to the highest three $p_T^{\rm jet}$ bins where the bias effects are negligible~\cite{STAR:2020xiv}.

From Fig.~\ref{fig:postparams_ATLAS_5020_0-10}, one can notice that the posterior distributions of $\beta$ and $\gamma$ are highly correlated, therefore it is more unambiguous to study the posterior of the averaged energy loss $\langle\Delta E\rangle$. We plot $\langle\Delta E\rangle$ at 65\% C.I.  as a function of the jet transverse momentum in Figs. \ref{fig:postElossATLAS_5020}, \ref{fig:postElossATLAS_2760}, and \ref{fig:postElossSTAR_200} for $A$+$A$ collisions with different centrality classes at three different colliding energies. The jet energy loss from each extraction is plotted only within the jet $p_T^{\rm jet}$ range of the dataset. It increases slowly with the jet transverse momentum. It is bigger in more central collisions and at higher colliding energies, implying that the jet energy loss increases with the medium density and system size or averaged propagation length. As we will show in the next section, the extracted parameter $\alpha$ for the jet energy loss fluctuation in Eq.~(\ref{eq:waa}) is approximately a constant within the accuracy of this analysis, independent of the colliding energy and centrality within the uncertainties at 65\% C.I.

\section{Momentum, density and system size dependence of jet energy loss}

The extraction of the averaged jet energy loss and its fluctuation from experimental data on $R_{AA}$ in the last section is  independently performed for each dataset with the assumed convolution form of the jet suppression factor $R_{AA}$ in  Eq.~(\ref{eq:raa}) without specifying the system size and medium density dependence of the jet energy loss. In this section, we will interpret the inferred jet energy loss distribution and discuss its dependence on jet momentum, initial medium density and system size or path-length of the collision system.

We first focus on the medium density and jet momentum dependence of the average jet energy loss $\langle\Delta E\rangle$. The averaged jet energy loss are extracted in different jet momentum range from systems with different beam energy and centrality. To factor out the system-size dependence from the discussion, we can first look at the collection of averaged jet energy loss in heavy-ion collisions at different beam energies but only in 0-10\% centrality class, where the averaged system sizes are expected to be similar. To be more precise, the averaged path-lengths in $A$+$A$ collisions at RHIC and LHC are slightly different due to several reasons, including 1) different mass number of the colliding nuclei ${}^{197}$Au versus ${}^{208}$Pb, 2) larger inelastic nucleon-nucleon cross-section $\sigma_{NN}^{\rm inel}\approx$70mb at $\sqrt{s}=5.02$ TeV than $\sigma_{NN}^{\rm inel}\approx 42$mb at $\sqrt{s}=200$ GeV, and 3) longer lifetime of the QGP fireballs at the LHC energies than at RHIC. However, these are all sub-leading effects and can be estimated through dynamic simulations. For our discussion here, we assume the average path length (including both QGP and hot hadronic matter) is proportional to the average system size, which is related to the average number of participant nucleons in each centrality class by $R_{\rm eff}\sim N_{\rm part}^{1/3}$.  The estimates of the system size for 0-10\% centrality at three beam energies are in good agreement as shown in Tab. \ref{tab:expdata}.

\begin{figure*}[ht!]
\centering
\includegraphics[height=.7\columnwidth]{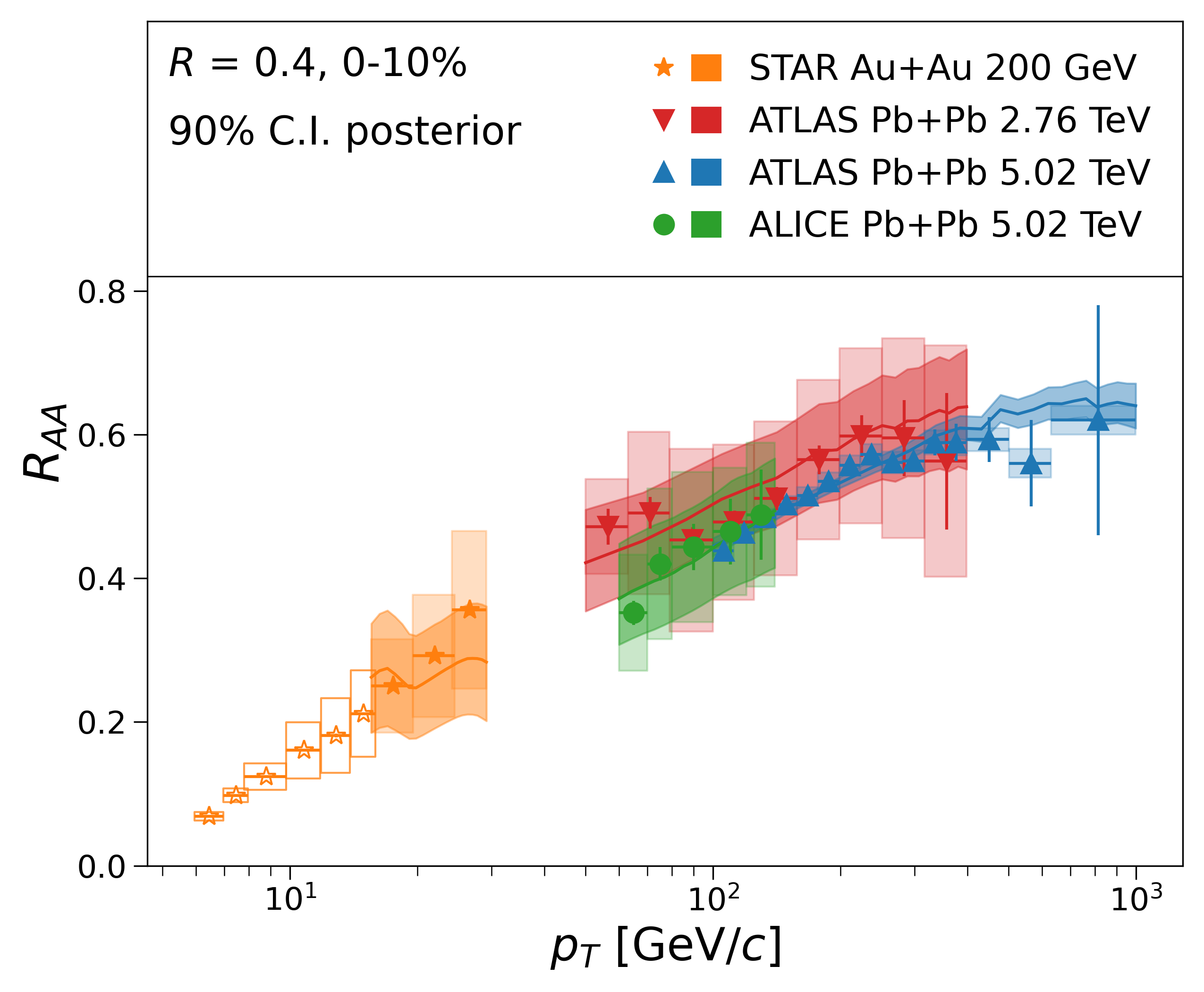}\quad\quad \includegraphics[height=.7\columnwidth]{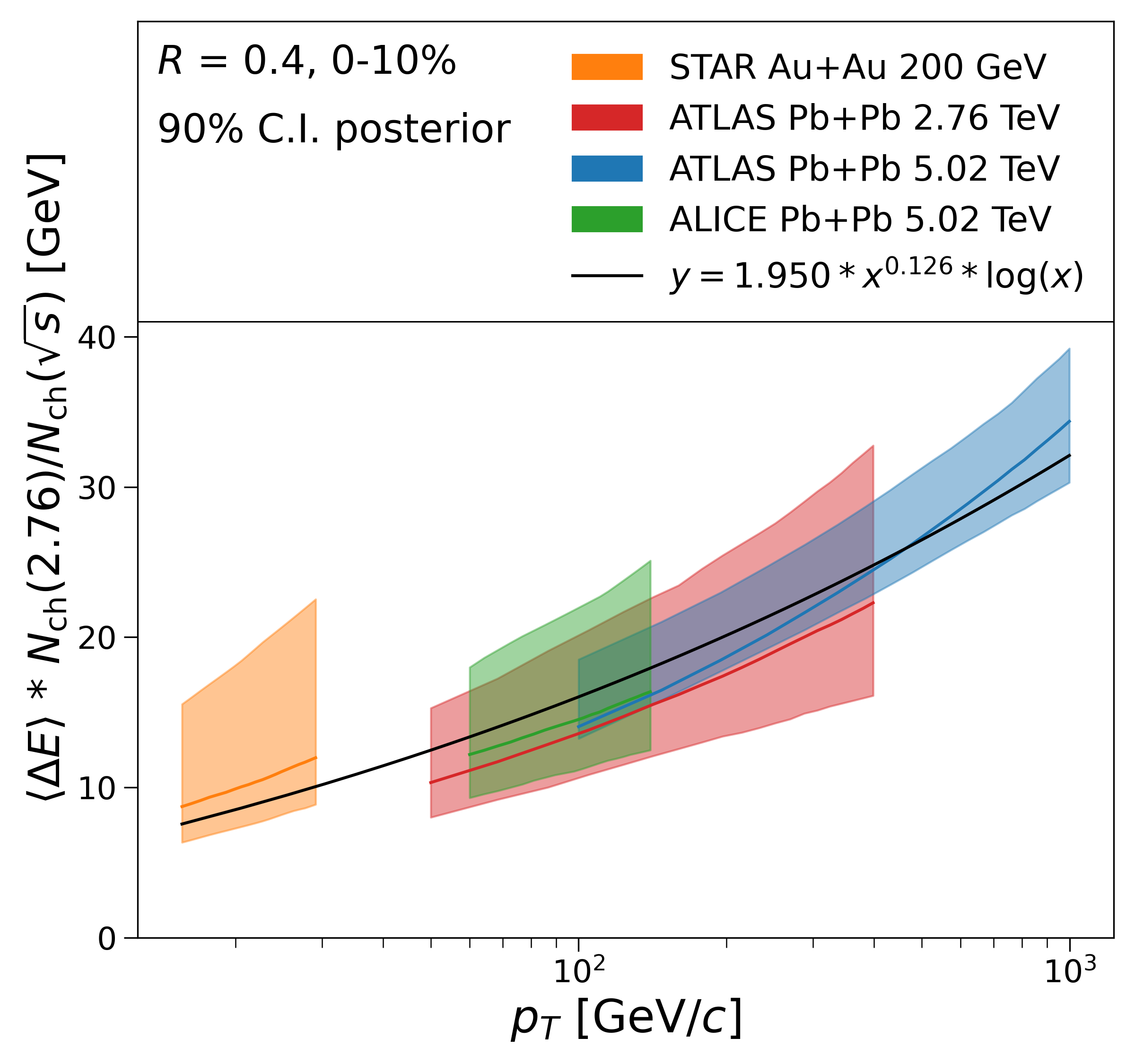}
 \caption{ (Left) The posterior single inclusive jet suppression factor $R_{AA}$ with $R=0.4$ at 90\% C.I. in 0-10\% central Pb+Pb collisions at $\sqrt{s}=2.76$ and 5.02 TeV and Au+Au collisions at $\sqrt{s}=200$ GeV, as compared to experimental data~\cite{ATLAS:2018gwx,ALICE:2019qyj,ATLAS:2018gwx,STAR:2020xiv}.     
    (Right) The corresponding posterior jet energy loss scaled by the  charge multiplicity density $(dN_{\rm ch}(\sqrt{s})/d\eta)/(dN_{\rm ch}(2.76 {\rm TeV})/d\eta)$. The black solid line $\sim p_T^{0.13}\ln p_T$ is a fit to the scaled posterior jet energy loss from all systems shown in this figure.
 }
 \label{fig:multiPostRAA}
\end{figure*}

In order to exam the colliding energy dependence of the jet energy loss distributions, we first plot in the left panel of Fig.~\ref{fig:multiPostRAA} the experimental data and posterior predictions at 90\% C.I. of the jet suppression factor $R_{AA}$ with jet cone size $R=0.4$ as a function of jet $p_T$ in 0-10\% central Au+Au at RHIC and Pb+Pb collisions at LHC energies. Across the range of colliding energies from RHIC and LHC, the suppression factor $R_{AA}$ seems to follow a common trend in its transverse momentum dependence in the respective kinematic ranges, increasing with $p_T$. The increase also seems to taper off at the respective upper bound of the kinematic region, corresponding to $x_T=2p_T/\sqrt{s}>0.3$.

Since the parton energy loss is proportional to the initial value of the jet transport coefficient or initial parton density~\cite{Baier:1996kr,Zakharov:1996fv,Wiedemann:2000za,Gyulassy:2000fs,Guo:2000nz,Wang:2001ifa,Majumder:2009ge} which in turn is proportional to the rapidity density $dN_{\rm ch}/d\eta$ of final charged multiplicity, we factor out such colliding energy loss dependence when compare the extracted jet energy loss from experimental data at different colliding energies. Shown in the right panel of Fig.~\ref{fig:multiPostRAA} is the extracted average jet energy loss scaled by the charged multiplicity density $dN_{\rm ch}(\sqrt{s})/dN_{\rm ch}(2.76 {\rm TeV})$ in central 0-10\% Au+Au collisions at $\sqrt{s}=200$ GeV and Pb+Pb collisions at $\sqrt{s}=2.76$ and 5.02 TeV \cite{PHENIX:2015tbb,ALICE:2010mlf, ALICE:2015juo}. The colored solid lines are the median energy loss and shaded bands are the 90\% C.I. We can see the scaled jet energy loss has a common momentum dependence which increases a little faster than a simple logarithmic dependence. 

Since the initial parton density can be estimated as proportional to $\rho\sim (dN_{\rm ch}/d\eta)/\pi R_{\rm eff}^2$ and the effective system size can be related to the average number of participant nucleons for a given centrality class $R_{\rm eff}\sim \langle N_{\rm part}\rangle^{1/3}$, we can parameterize the jet energy loss as,
\begin{equation}
    \langle \Delta E \rangle =\frac{dN_{\rm ch}(\sqrt{s})}{d\eta}\frac{1}{\langle N_{\rm part} \rangle^{2/3}} f\left(\langle N_{\rm part}\rangle^{1/3} \right) g(p_T).
\end{equation}
where $f\left(\langle N_{\rm part}\rangle^{1/3}\right)$ and $g(p_T)$ are now assumed to be universal functions for all collision systems and centralities.
Fitting to the momentum dependence of the extracted jet energy loss in the right panel of Fig.~\ref{fig:multiPostRAA} for central 0-10\% Au+Au and Pb+Pb collisions, we find $g(p_T)\approx p_T^{0.13}\ln p_T$ which is slightly stronger than a logarithmic dependence. This is consistent with the earlier Bayesian analysis~\cite{He:2018gks}. The $p_T$ ranges of available experimental data on single inclusive jet spectra with other centralities are smaller than the most central collisions in Fig.~\ref{fig:multiPostRAA}. The extracted $p_T$ dependence of the jet energy loss for these semi-central and semi-peripheral collisions is consistent with the above functional form.

\begin{figure*}[ht!]
\centering
\includegraphics[height=.8\columnwidth]{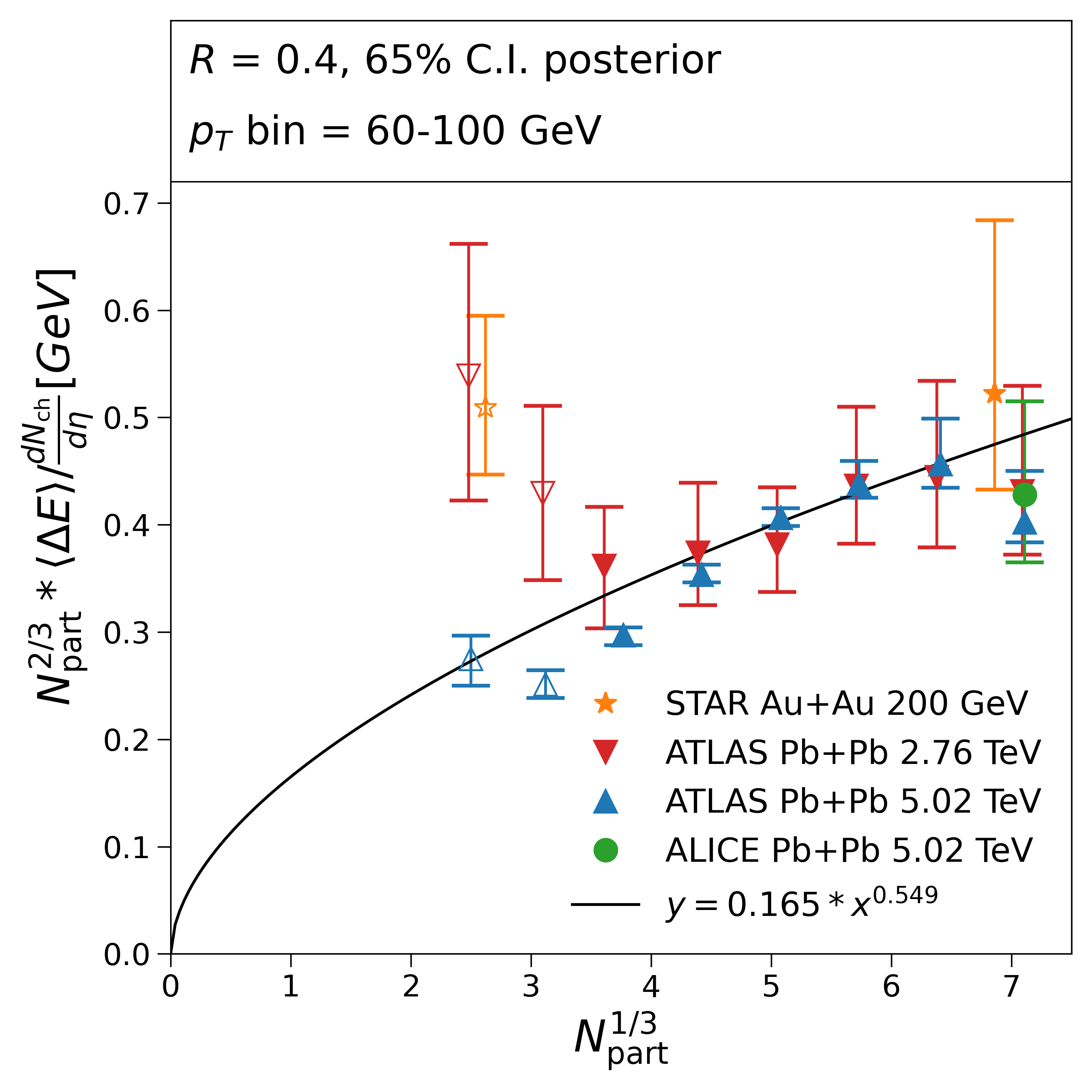} \quad \quad \includegraphics[height=.8\columnwidth]{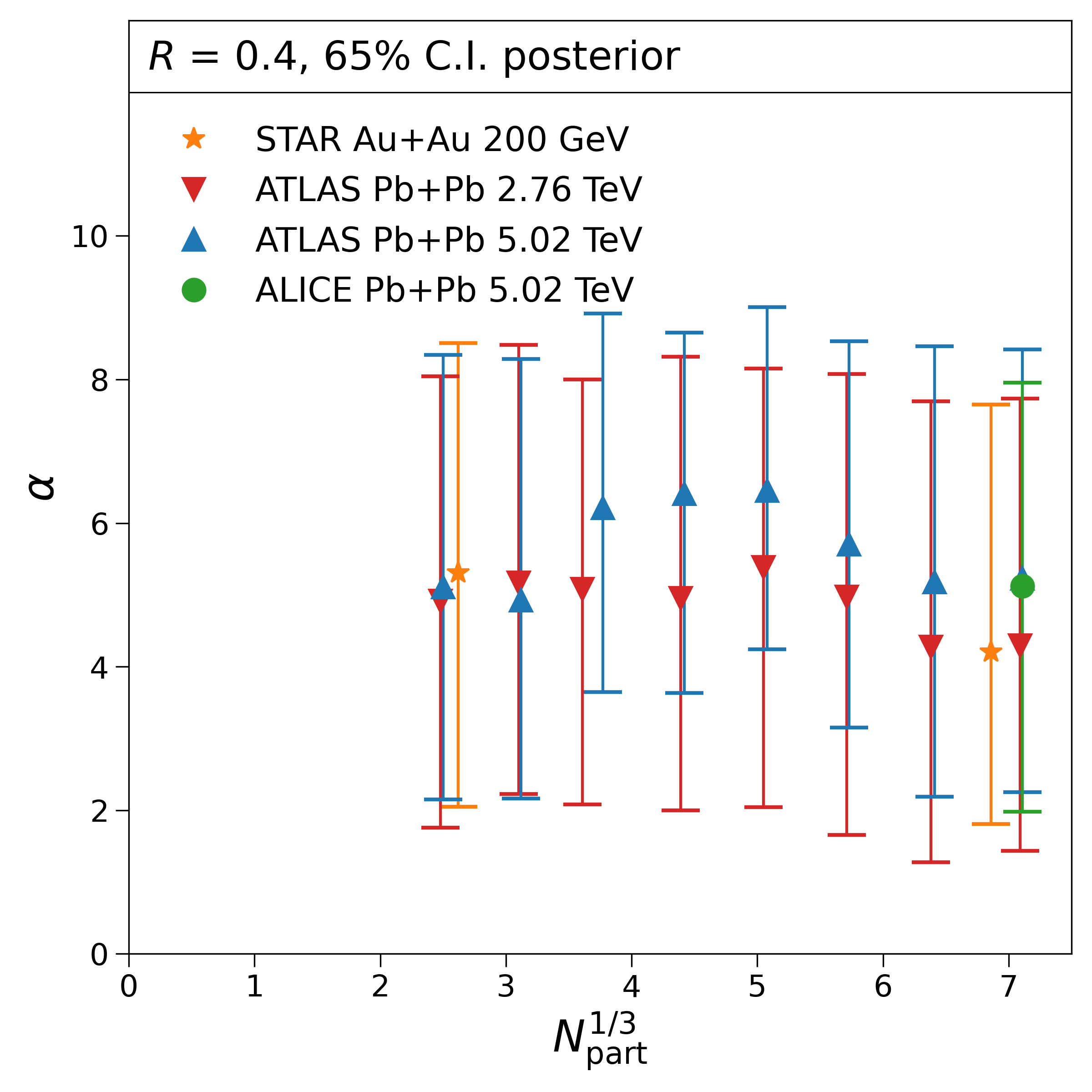}
 \caption{The posterior (Left) average jet energy loss scaled by the initial medium density  $(dN_{\rm ch}/d\eta)/\langle N_{\rm part}\rangle^{2/3}$ and (Right) the parameter $\alpha$ for jet energy loss fluctuations as a function of $\langle N_{\rm part}\rangle^{1/3}.$
The solid line is a power-law fit $(\langle N_{\rm part}\rangle^{1/3})^{0.557}$ excluding the peripheral collisions with $\langle N_{\rm part}\rangle^{1/3}<3.3$.
 }
 \label{fig:ElossNpart}
\end{figure*}

To find out the system size dependence of the extracted jet energy loss, we plot in the left panel of Fig.~\ref{fig:ElossNpart} the scaled jet energy loss $\langle N_{\rm part}\rangle^{2/3} \langle \Delta E \rangle /(dN_{\rm ch}/d\eta)$ as a function of $\langle N_{\rm part}\rangle^{1/3}$ for final jet $p_T=60-100$ GeV/$c$ in Au+Au collisions at RHIC and Pb+Pb collisions at LHC energies with different centrality classes. The values of $\langle N_{\rm part}\rangle^{1/3}$ from Glauber model are listed in Tab. \ref{tab:expdata}. Excluding the extracted values in very (60-80\%) peripheral collisions where the Bayesian fitting does not do well,  the extracted jet energy loss has an approximate $f(\langle N_{\rm part}\rangle^{1/3}) \sim (\langle N_{\rm part}\rangle^{1/3})^{0.59}$ dependence on the system size. This is quite different from the approximate linear dependence on the system size for the energy loss of a single parton in an expanding system \cite{Gyulassy:2001kr}, which is needed to explain system size dependence of the suppression of single inclusive hadron spectra \cite{Arleo:2022shs}. Jet energy loss, however, is not proportional to energy loss of individual shower partons. Energy carried by radiated gluons and medium response within the jet cone is recovered by the jet, reducing the jet energy loss. This will both lead to the weaker system size dependence as shown by the LBT model simulations \cite{He:2018xjv,He:2018gks}.

We also extract the parameters $\alpha$ for the jet energy loss fluctuations in Eq.~(\ref{eq:waa}) and they are shown in the right panel of Fig.~\ref{fig:ElossNpart} as a function of $\langle N_{\rm part}\rangle^{1/3}$.  Within the larger errors at 65\% C.I., the jet energy loss fluctuation parameter $\alpha\approx 5.0 \pm 3$ does not depend on the system size and the initial parton density (or colliding energy).

\section{Summary}

Employing Bayesian inference, we have analyzed the world data on the nuclear suppression of single inclusive jet spectra in heavy-ion collisions at both RHIC and LHC energies with a wide selection of centrality classes. We have extracted the average jet energy loss as a function of the jet transverse momentum for each centrality class of collisions at each colliding energy and the jet energy loss fluctuations. We found that the extracted average jet energy loss scales with the initial parton density $\rho \propto (dN_{\rm ch}/d\eta)/\langle N_{\rm part}\rangle^{2/3}$. The average jet energy loss scaled by the initial parton density has a jet momentum dependence $\langle \Delta E\rangle/\rho \propto p_T^{0.13} \ln p_T$ that is slightly stronger than a logarithmic form. It has a system size dependence $\langle \Delta E\rangle/\rho \propto (\langle N_{\rm part}\rangle^{1/3})^{0.59}$. This behavior of jet energy loss is different from that of a single energetic parton. Such a difference can be attributed to energy loss carried by the radiated gluons and medium response outside the jet cone. 

\section*{Acknowledgements}

This work is supported by the U.S. Department of Energy, Office of Science, Office of Nuclear Physics under Contracts No. DE-AC02-05CH11231 and No. 89233218CNA000001, and by NSF under Grant No. OAC-2004571 within the X-SCAPE Collaboration. W.K. is also supported by the Laboratory Directed Research and Development Program at LANL.

\bibliography{baysian-eloss.bib}

\end{document}